\documentclass[preprint,aps]{revtex4}

\begin{document}
%\draft
\newcommand{\pp}[1]{\phantom{#1}}
\newcommand{\be}{\begin{eqnarray}}
\newcommand{\ee}{\end{eqnarray}}
\newcommand{\ve}{\varepsilon}
\newcommand{\vs}{\varsigma}
\newcommand{\Tr}{{\,\rm Tr\,}}
\newcommand{\pol}{\frac{1}{2}}
\newcommand{\ba}{\begin{array}}
\newcommand{\ea}{\end{array}}
\newcommand{\bear}{\begin{eqnarray}}
\newcommand{\eear}{\end{eqnarray}}
\newtheorem{lemma}{Lemma}
\newtheorem{theorem}{Theorem}
\newtheorem{proposition}{Proposition}

\title{
Construction of exact solutions of Bloch-Maxwell equation based on
Darboux transformation}
\author{Maciej Kuna}
\altaffiliation{Departement Natuurkunde, Universiteit Antwerpen UIA, B2610 Antwerpen, Belgium\\
and\\
Wydzia{\l} Fizyki Technicznej i Matematyki Stosowanej,
Politechnika Gda\'nska, 80-952 Gda\'nsk, Poland}
\email{maciek@mifgate.mif.pg.gda.pl}

\begin{abstract}
A new strategy, using Darboux transformations, of finding
self-switching solutions of $i\dot \rho=[H,f(\rho)]$ is
introduced. Unlike the previous ones, working for any $f$ but for
Hamiltonians whose spectrum contains at least three equally spaced
eigenvalues, the strategy does not impose any restriction on the
discrete part of spectrum of $H$. The strategy is applied to the
Bloch-Maxwell system.
\end{abstract}
\maketitle

%\narrowtext

%%%%%%%%%%%%%%%%%%%%%%%%%%%%%%%%%%%%%%%%%%%%%%%%%%%%%%%%%%%%%%%%%%%%%%%
%%%%%%%%%%%%%%%%%%%%%%%%%%%%%%%%%%%%%%%%%%%%%%%%%%%%%%%%%%%%%
\section{Introduction}

The nonlinear von Neumann equation (NvNE)
\cite{MC1997,LC,MCJN,UCKL} \be i\dot \rho=[H,f(\rho)]\label{fvN}
\ee was recently shown to be relevant for two apparently
disconnected fields of research: Optical solitons \cite{CDSW} and
chemical kinetics \cite{ACS}. In both cases we are interested in
solutions $\rho$ that have a form of $n\times n$ matrices. In
optical applications the dimension $n$ is related to $n$-level
atoms. For chemical-type kinetics the  kinetic variables are
related to real and imaginary parts of off-diagonal matrix
elements of $\rho$ evaluated in the basis of eigenvectors of $H$.
Since any chemical kinetics involves a finite number of
substrates, the dimension $n$ must also be finite. It has to be
stressed, however, that a class of infinite-dimensional solutions
of (\ref{fvN}) was also found, albeit for a rather artificial
example, in \cite{UCKL}. Moreover, as shown in \cite{CCU} the NvNE
contains as special cases many known lattice equations that also
lead to infinitely-dimensional solutions.

The soliton-type solutions of (\ref{fvN}) were first found in
\cite{LC} for $f(\rho)=\rho^2$ and a special class of Hamiltonians
$H$. The Hamiltonians $H$ could be characterized as those that
contained at least three equally spaced eigenvalues in discrete
parts of their spectra. There are, of course, many physical
examples of such systems, the most notable cases including
harmonic oscillators, quantum fields or spin systems. However, the
problem that was not solved until now was how to construct
solutions for $H$ that has an {\it arbitrary\/} spectrum. In the
present paper we show how to deal with any discrete spectrum and
illustrate the technique on a harmonic oscillator (HO) and a
hydrogen-type atom (HA).

The method we employ is based on a Darboux transformation. The
main advantage of the technique is that it allows to construct
solutions that are practically impossible to find by other means,
especially if $n$ is large or infinite. The disadvantage lies in
the need of starting with a known solution that, in addition,
guarantees nontriviality of the solution generated by the Darboux
transformation.

We work out explicitly two important cases: $n=2$ and $n=3$. The
cases are important for two main reasons. Firstly, the two
dimensional case is easy to handle and allows for a seed solution
having a sufficient number of arbitrary parameters, no matter what
spectrum of $H$ one encounters. Secondly, once we have
two-dimensional seed solutions, we can use them to construct
highly nontrivial ``self-scattering" solutions involving any even
$n$. If $n$ is odd we can employ a combination of two- and
three-dimensional seed solutions.

In the present paper we restrict the analysis to the essential
building blocks with $n=2$ and $n=3$. But even these cases,
especially $n=3$, are interesting in themselves. For quantum
optics we can apply them to three-level atoms and the result is a
dynamics analogous to the celebrated ``sech" soliton \cite{CDSW}.
In application to chemical kinetics the case results in an
Oregonator-type dynamics with four substrates \cite{ACS}.

In both cases it is sufficient to work with quadratic
nonlinearities. Kinetic autocatalytic-type equations are found
immediately by considering separately real and imaginary parts of
matrix elements $\langle k|\rho|l\rangle$ where $|k\rangle$,
$|l\rangle$ are eigenvectors of $H$ normalized to $\langle
k|l\rangle=\delta_{kl}$.

In order to relate the dynamics to $n$-level atoms one makes the
trick introduced in \cite{CDSW}. Take a solution $\rho$ of $ i\dot
\rho=[H,\rho^2] $ and employ the algebraic formula
$[H,\rho^2]=[H\rho+\rho H,\rho]$. For a given solution $\rho$
rewrite $H\rho+\rho H$ as $H'+V$ where $H'$ has the same form as
$H$ but with modified parameters. Then $V$ is a time-dependent
interaction term that has a form $- d \cdot E(t, 0)$ where $ E(t,
0)$ is a field typical of an optical soliton at the origin $ x=
0$. As pointed out in \cite{CDSW} the construction is analogous to
SUSY quantum mechanics since we have in addition the Darboux
transformation $\rho\mapsto\rho_1$ that allows for the
transformation $V\mapsto V_1$. We leave these details for a later
work but in the present paper concentrate on technicalities
leading to the solutions for arbitrary $H$.

\section{Lax pair and Darboux transformation}

Following \cite{LC,UCKL,KCL} one starts with three overdetermined
linear systems (Lax pairs) \be z_k |\varphi_k \rangle &=&
(\rho -\mu_k H)|\varphi_k\rangle,\label{3}\\
i|\dot\varphi_k\rangle &=& \frac{1}{\mu_k}f(\rho
)|\varphi_k\rangle, \label{4}\ee where $\rho$, $H$ are
self-adjoint operators on some Hilbert space ${\cal H}$, $\mu_k$,
$z_k$ are complex numbers, $f$ is real function and
$|\varphi_k\rangle$ are elements of ${\cal H}$ for $k = 1,2,3$. If
we fix an operator $H$ to be time-independent then there exist an
operator $\rho $ and a vector $|\varphi_k\rangle$ that fulfil the
Lax pair only if $ i\dot\rho |\varphi_k\rangle = [H,f(\rho
)]|\varphi_k\rangle$. The basic theorem \cite{UCKL} is the
following:

\medskip\noindent
{\bf Theorem.} Assume $|\varphi_k\rangle$ for $k = 1,2,3$ are
solutions of (\ref{3}, \ref{4}) respectively and
$|\psi[1]\rangle$, $\rho[1]$,  are defined by \be
|\psi[1]\rangle&=&\Big( 1 +
\frac{\mu_2-\bar\mu_3}{\bar\mu_3-\mu_1}\texttt{P}^*\Big)|\varphi_1\rangle
,\label{_1_a}\\
\rho[1] &=& \Big( 1+\frac{\mu_3
-\bar\mu_2}{\bar\mu_2}\texttt{P}\Big) \rho \Big( 1+\frac{\bar\mu_2
-\mu_3}{\mu_3}\texttt{P}\Big), \label{_1_b}
\\
f(\rho[1]) &=& \Big( 1+\frac{\mu_3
-\bar\mu_2}{\bar\mu_2}\texttt{P}\Big) f(\rho) \Big(
1+\frac{\bar\mu_2 -\mu_3}{\mu_3}\texttt{P}\Big), \label{_1_c}
\\
\texttt{P} &=& \frac{|\varphi_1 \rangle\langle\varphi_2
|}{\langle\varphi_2 |\varphi_1\rangle} .\label{P} \ee Then \be z_1
|\psi[1]\rangle &=&
(\rho[1]^* -\mu_k H)|\psi[1]\rangle,\\
i|\dot\psi[1]\rangle &=& \frac{1}{\mu_1}f(\rho
[1])^*|\psi[1]\rangle, \ee

If we put $\mu_3=\mu_2$ and $|\varphi_1\rangle =|\varphi_2
\rangle$ then $\rho[1]$ is also a density matrix and spectra of
$\rho$ and $\rho[1]$ are identical. In this case (\ref{_1_c}) is
simply equivalent with definition of function $f$ of operator
$\rho[1]$ given by the spectral theorem. We can proof \cite{NK}
that if $\rho$ is a solution of (\ref{fvN}) then $\rho[1]$ is also
a solution of (\ref{fvN}).

\section {Two-dimensional case}

We start with a Hamiltonian having discrete spectrum: $H = \sum_n
h_n Q_n$, where $Q_n$ and $h_n$ are spectral projectors and
eigenvalues of $H$ respectively. Next we take a density matrix
$\rho$ with two spectral projectors $P_1$ and $P_2$ which satisfy
the following conditions: $[P_1 , Q_n ]= [P_2, Q_n ]= 0$ for all
$n \neq 1,2$ Therefore ${\bf P} = P_1 + P_2 = Q_1 + Q_2 $ is a
projector on the two-dimensional common eigensubspace. Let us
define $\rho_1 = {\bf P} \rho {\bf P}$, $H_1 = {\bf P} H {\bf P}$,
${\rho}'= (1 - {\bf P}) \rho(1 - {\bf P})$ and $H' = (1 - {\bf P})
H (1 - {\bf P})$. Then equation $i\dot{\rho} = [H, f(\rho ) ]$
separates into two pieces $i\dot{\rho_1} = [H_1 , f(\rho_1 ) ]$
and $i\dot{{\rho}'} = [H', f({\rho}' ) ]$. By using the relation
$[{\bf P},H_1 ]=0$
  and properties of such evolution $\dot{\bf
P}=0$ the first equation can be
 described by spectral projectors:
\begin{eqnarray}
i\dot{\rho_1} &=& i\dot{(\lambda_1 P_1 + \lambda_2 P_2 )} =
 i(\lambda_1 - \lambda_2 )\dot{P_1}\cr
 [H_1 , f(\rho_1 ) ] &=&
[(f(\lambda_1 ) - f(\lambda_2 )) H_1 , P_1 ] ,
\end{eqnarray}
so
$$i\dot{P_1} =[\alpha (\lambda ) H_1 , P_1 ] $$
and similarly:
$$i\dot{P_2} =[\alpha (\lambda )  H_1 , P_2 ], $$
where  $\alpha (\lambda ) = \frac{f(\lambda_1 ) - f(\lambda_2
)}{\lambda_1 - \lambda_2 } $. Solutions of these equations are the
following:
$${P_i}(t) = e^{-i\alpha (\lambda )H_1 t} {P_i}(0) e^{i\alpha (\lambda )H_1 t}$$
for $i = 1,2$  and we have the general solution of two-dimensional
 problem:
$${\rho_1 }(t) = e^{-i\alpha (\lambda )H_1 t} {\rho_1 }(0) e^{i\alpha (\lambda )H_1 t}.$$

We want to use it in further constructions, therefore we have to
examine its properties. Let us assume that $\rho_1$ together with
$H_1$ and fixed $z_{\mu }$, $\mu$ and $|\varphi_1 \rangle $ fulfil
the Lax pair on ${\bf P}{\cal H}$:
\begin{eqnarray}
z_{\mu }|\varphi_1 \rangle  &=& (\rho_1 - \mu H_1 )|\varphi_1
\rangle \cr i\dot{|\varphi_1 \rangle } &=& \frac{1}{\mu
}f_c(\rho_1 )|\varphi_1 \rangle \label{lp2} ,
\end{eqnarray}
where $f_c(\rho_1 ) = f(\rho_1 ) + c{\bf P}$.

Put $|\psi_1 \rangle  = e^{i\alpha (\lambda )H_1 t}|\varphi_1
\rangle $, then

\begin{eqnarray}
z_{\mu }|\psi_1 \rangle  &=& ({\rho_1 }(0)- \mu H_1 )|\psi_1
\rangle \cr i\dot{|\psi_1 \rangle } &=& \{ \frac{1}{\mu
}f_c({\rho_1 }(0)) - \alpha (\lambda )H_1 \} |\psi_1 \rangle .
\end{eqnarray}
So $|\psi_1 (t)\rangle  = e^{-i\{\frac{1}{\mu }f_c({\rho_1 }(0)) -
\alpha (\lambda )H_1 \} t}|\psi_1 (0)\rangle $. It is easy to
check that the generator of time evolution of $|\psi_1 \rangle $
commutes with ${\rho_1 }(0)- \mu H_1$.

Taking this into account we obtain
\begin{eqnarray}
z_{\mu }|\psi_1 (0)\rangle  &=& ({\rho_1 }(0)- \mu H_1 )|\psi_1
(0)\rangle \cr |\varphi_1 (t)\rangle  &=& e^{-i\alpha (\lambda
)H_1 t} e^{-i\{\frac{1}{\mu }f_c({\rho_1 }(0)) - \alpha (\lambda
)H_1 \} t}|\psi_1 (0)\rangle .
\end{eqnarray}
We can conclude that if $\rho_1$ is a solution of the
two-dimensional problem then it is enough to consider only the Lax
pair for time $t=0$.
\section {Three-dimensional case}

We construct a density matrix $\rho$ using only three
eigenprojectors: $P_1$, $P_2$ and third spectral projector is
equal to $Q_3$. So ${\rho}'= \lambda_3 P_3 = \lambda_3 Q_3$.
Denote by $|\omega_i (0)\rangle $ eigenvectors of ${\rho }(0)$ ,
so $ P_i (0) =|\omega_i (0)\rangle \langle \omega_i (0)|$ and
denote by $ P_{i,j} (0) =|\omega_i (0)\rangle \langle \omega_j
(0)|$. Now we want to add to the Lax pair (\ref{lp2}) a component
on this one-dimensional space in such a way that together they
will be the three-dimensional Lax pair with slightly modified $f$.
For this case we can also restrict to initial conditions and for
given $z_{\mu }$, $\mu$ and $h_3$ we can choose $\lambda_3$
satisfying the Lax pair:
\begin{eqnarray}
z_{\mu } |\omega_3 (0)\rangle  &=& (\lambda_3 - \mu h_3 )|\omega_3
(0)\rangle \cr |\omega_3 (t)\rangle  &=& |\omega_3 (0)\rangle .
\end{eqnarray}
Put $|\varphi (t)\rangle  =\gamma_1 |\varphi_1 (t)\rangle  +
\gamma_3|\omega_3 (0)\rangle $ for arbitrary $\gamma_1 , \gamma_3
\in {\cal C}$. Then we have the three-dimensional Lax pair:
\begin{eqnarray}
z_{\mu }|\varphi \rangle  &=& (\rho - \mu H )|\varphi \rangle \cr i\dot{|\varphi
\rangle } &=& \frac{1}{\mu }(f(\rho ) - f(\lambda_3 ))|\varphi \rangle  \equiv \frac{1}{\mu }f'(\rho )|\varphi \rangle
\end{eqnarray}
and we have to consider only
\begin{eqnarray}
z_{\mu }|\psi (0)\rangle  &=& ({\rho }(0)- \mu H )|\psi (0)\rangle
\cr
 |\varphi(t)\rangle  &=& e^{-i\alpha (\lambda )H t} e^{-i\{\frac{1}{\mu
}f'({\rho }(0)) - \alpha (\lambda )H \} t}|\psi (0)\rangle \equiv
U|\psi (0)\rangle  \label{A}
\end{eqnarray}
with  $|\psi \rangle  = e^{i\alpha (\lambda )H t}|\varphi \rangle
$.

Now we have the Lax pair expressed only by initial conditions and
can construct the operator $\texttt{P}$ which defines the Darboux
transformation (see \ref{P}):
\begin{eqnarray}
\texttt{P} &=& \frac{ |\varphi (t)\rangle  \langle \varphi
(t)|}{\langle \varphi (t)|\varphi (t)\rangle } = \frac{ U|\psi
(0)\rangle  \langle \psi (0)|e^{i\{\frac{1}{\overline{\mu
}}f'({\rho }(0)) - \alpha (\lambda )H \} t}e^{i\alpha (\lambda )H
t}}{\langle \varphi (t)|\varphi (t)\rangle }\cr &=& e^{-i\alpha
(\lambda )H t}\texttt{P}_{int} e^{i\alpha (\lambda )H t} \equiv V
\texttt{P}_{int} V^{*},
\end{eqnarray}
where
$$\texttt{P}_{int} = \frac{ |\psi (t)\rangle  \langle \psi (t)|}{\langle \varphi (t)|\varphi
(t)\rangle } = \frac{ |\psi (t)\rangle  \langle \psi (t)|}{\langle
\psi (t)|\psi (t)\rangle }.$$ Then
\begin{eqnarray}
\rho \left[ 1\right](t) &=&  ( 1+ \frac{\mu - \overline{\mu
}}{\overline{\mu }}\texttt{P}) \rho (t)( 1+ \frac{\overline{\mu }
- \mu }{\mu }\texttt{P})\cr &=&V ( 1+ \frac{\mu - \overline{\mu
}}{\overline{\mu }}\texttt{P}_{int} ) {\rho }(0) ( 1+
\frac{\overline{\mu } - \mu }{\mu }\texttt{P}_{int} ) V^{*} = V
\rho_{int} \left[ 1\right](t)V^{*}. \label{sandw}
\end{eqnarray}
From the first equation of Lax pair (\ref{3}) we have
 $$H |\psi \rangle  = \frac{1}{\mu
}({\rho }(0) - z_{\mu })|\psi \rangle $$ so
$$ (\frac{1}{\mu
}f'({\rho }(0)) - \alpha (\lambda )H )|\psi \rangle
 = \frac{1}{\mu }(f'({\rho }(0)) - \alpha (\lambda )[{\rho}(0) - z_{\mu }])|\psi \rangle $$
and the second equation of Lax pair reads
$$i\dot{|\psi \rangle }= \frac{1}{\mu }(f'({\rho }(0)) - \alpha (\lambda )[{\rho }(0) - z_{\mu }])|\psi \rangle .$$
It can be written as
\begin{eqnarray}
|\psi (t)\rangle  &=& e^{ \frac{-i}{\mu }\{f'({\rho }(0)) - \alpha
(\lambda )[{\rho }(0) - z_{\mu }] \} t}|\psi (0)\rangle  =
  e^{
\frac{-i}{\mu }\{ \sum_{i=1}^3(f'(\lambda_i ) - \alpha (\lambda
)[\lambda_i  - z_{\mu }])P_i \} t}|\psi (0)\rangle \cr &\equiv&
e^{ \frac{-i}{\mu }\{ \sum_{i=1}^3 \beta(\lambda_i )P_i \} t}|\psi
(0)\rangle ,
\end{eqnarray}
where $\beta(\lambda_i ) = f'(\lambda_i ) - \alpha (\lambda
)[\lambda_i  - z_{\mu }]$. Inserting this into the definition of $\texttt{P}_{int}$
we obtain
\begin{eqnarray}
\texttt{P}_{int} &=& \frac{ |\psi (t)\rangle  \langle \psi
(t)|}{\langle \psi (t)|\psi (t)\rangle } =\frac{1}{F(t)} \times
\{\sum_{i,j=1}^3 e^{-i (\frac{\beta(\lambda_i )}{\mu } -
\frac{\overline{\beta(\lambda_j )}}{\overline{\mu }})}P_i (0)
|\psi (0)\rangle  \langle \psi (0)|P_j (0) \} \cr &=&
\frac{\sum_{i,j=1}^3 a_{ij} e^{-ib_{ij}t }P_{ij} (0)} {F(t)},
\end{eqnarray}
where  $a_{ij} =\langle \omega_i (0)|\psi (0)\rangle
\overline{\langle \psi (0)|\omega_j (0)\rangle }$, $b_{ij} =
\frac{\beta(\lambda_i )}{\mu } - \frac{\overline{\beta(\lambda_j
)}}{\overline{\mu }}$. Since $b_{11} = b_{22} = b_{12} = b_{21}
\equiv b$ and $\beta(\lambda_3 ) = - \alpha (\lambda )\mu h_3$
then $b_{33} = 0$. So $F(t)= \sum_{i=1}^3 a_{ii} e^{-ib_{ii}t } =
e^{-ibt}(a_{11} + a_{22}) + a_{33}$. If $\gamma_3 = 0$ then the
projector $\texttt{P}_{int}$ does not depend on time, so we need
at least a three dimensional space to have a nontrivial Darboux
transformation. We are in position to give the form of $\rho_{int}
\left[ 1\right](t)$ in terms of spectral projectors of ${\rho
}(0)$:
\begin{eqnarray}
\rho_{int} \left[ 1\right](t) &=&  ( 1+ \frac{\mu - \overline{\mu
}}{\overline{\mu }}\texttt{P}_{int}) \rho (0)( 1+
\frac{\overline{\mu } - \mu }{\mu }\texttt{P}_{int} )\cr &=&
 \rho (0) + \frac{(\mu
- \overline{\mu })}{F(t)^2|\mu |^2} \{\sum_{i=1}^3 c_i P_i(0)+
\sum_{i\neq j} c_{ij} P_{ij}(0)\},
\end{eqnarray}
where $c_i = (\mu - \overline{\mu })
a_{ii}e^{-ib_{ii}t}[(\lambda_i - \lambda_j)a_{jj}e^{-ib_{jj}t} +
(\lambda_1 - \lambda_k)a_{kk}e^{-ib_{kk}t}]$ and $c_{ij} =
a_{ij}e^{-ib_{ij}t}[(\lambda_j - \lambda_i)(\mu
a_{ii}e^{-ib_{ii}t} + \overline{\mu }a_{jj}e^{-ib_{jj}t}) +
((\overline{\mu } - \mu )\lambda_k + \mu \lambda_j -
\overline{\mu}\lambda_i )a_{kk}e^{-ib_{kk}t} ]$ and here $j$, $k$
in the first term and $k$ in the second term mean the remaining indexes.

\section {Models}

We want to see whether we can find appropriate $\mu$ and a density
matrix $\rho$ for a construction of ``self-scattering" solution
for any Hamiltonian $H$ with a discrete part of spectrum and
nonlinearity $f$. We use the spectral representation of $H$ and
take any three of its eigenvectors:
\begin {equation}
H = \left(\begin{array}{ccc} h_1  & 0  &0 \cr 0 &h_2 &0\cr
 0 & 0 &h_3\cr
\end{array}\right)
\end {equation}
We can decompose $H$ in two pieces in two different ways:
\begin {equation}
H = \left(\begin{array}{ccc}
 \frac{1}{2 }(h_1 - h_2) & 0  &0 \cr
 0 & -\frac{1}{2 }(h_1 - h_2) &0\cr
 0 & 0 & h_3 - \frac{1}{2 }(h_1 + h_2)\cr
\end{array}\right) +
[h_1 -\frac{1}{2 }(h_1 - h_2)]I
\end {equation}
for $h_3 - \frac{1}{2 }(h_1 + h_2) \neq 0$ and
\begin {equation}
H = \left(\begin{array}{ccc}
B & 0  &0 \cr
 0 & 3B &0\cr
 0 & 0 &2B \cr
\end{array}\right) +
AI
\end {equation}
for equally-spaced spectrum. Clearly, if the density matrix $\rho$
is a solution of the equation $i\dot{\rho} = [H, f(\rho ) ]$, then
it also satisfies the equation with $H'= H + C I$, so we can
consider Hamiltonians with special forms. In the first case, for
the operator
\begin {equation}
\rho - \mu H = \left(\begin{array}{ccc}
 \varrho_1 -  \frac{\mu}{2 }(h_1 - h_2)& \varrho &0 \cr
 \overline{\varrho } &  \varrho_2 + \frac{\mu}{2 }(h_1 - h_2) &0\cr
 0 & 0 & \varrho_3 - \mu [h_3 - \frac{1}{2 }(h_1 + h_2)]\cr
\end{array}\right)
\end {equation}
we can choose $ \varrho_1 - \varrho_2 = \alpha (h_1 - h_2 )$,
${\beta }^2(h_1 - h_2 )^2 \rangle  4|\varrho |^2$, $|\varrho |^2 = {\beta
}^2[h_3(h_1 + h_2 - h_3) - h_1 h_2]$ and $ \varrho_1 + \varrho_2 =
2\varrho_3 - 2\alpha h_3 + \alpha (h_1 + h_2 )$, where $\mu =
\alpha + i\beta$ to have an appropriate eigenvector. The compatibility of
the third assumption is satisfied if we arrange eigenvalues of $H$ in
such a way that $h_3$ is a number between $h_1$ and  $h_2$. Under
these assumptions we can take $\varrho_1 = \alpha (h_1 - h_3 ) +
\varrho_3$, $\varrho_2 = \alpha (h_2 - h_3 ) + \varrho_3$ and
$|\varrho |^2 = {\beta }^2[h_3(h_1 + h_2 - h_3) - h_1 h_2]$. To
get $\rho$ positive we need $\varrho_i
\rangle  0$ and $\varrho_1\varrho_2 \rangle  |\varrho |^2$. We can easy control
these conditions by choosing an appropriate $\varrho_3$, $\alpha$ and
$\beta$. In order to have $\rho$ as a density matrix we have to
put $\alpha = \frac{1 - 3\varrho_3}{h_1 + h_2 - 2h_3}$. We will
latter discuss this condition. For such a specification $\rho -
\mu H$ has the eigenvalue $z_{\mu} =\varrho_3 + \frac{1}{2 }(1 - 3\varrho_3) -
i\beta [h_3 - \frac{1}{2 }(h_1 + h_2)]$.

In the second case:
\begin {equation}
\rho' - \mu H = \left(\begin{array}{ccc}
 \varrho'_1 - {\mu}B& \varrho' &0 \cr
 \overline{\varrho' } &  \varrho'_2 - 3{\mu}B &0\cr
 0 & 0 & \varrho'_3 - 2{\mu}B\cr
\end{array}\right)
\end {equation}
has an appropriate two-dimensional eigenspace if,  for example, we
take $\varrho'_1 - \varrho'_2 = -2\alpha B$ and $|\varrho' |^2 =
{\beta }^2 B^2$. To obtain density matrix we can put $\varrho'_3 =
\frac{1}{3}$, $\varrho'_1 = \frac{1}{3} - \alpha B$, $\varrho'_2 =
\frac{1}{3} + \alpha B$, $|\varrho' |^2 = {\beta }^2 B^2$ and
$|\mu |^2 B^2 \langle  \frac{1}{3}$. Then $z_{\mu} = \frac{1}{3 }
- 2{\mu}B$.

Let us choose a hydrogen-like atom Hamitonian (HA) as an example of an inhomogeneous
spectrum and a harmonic oscillator (HO) for the homogeneous one. We
choose neighbouring levels for simplicity: $h_{1}^{(HA)} =
\frac{-B^{(HA)}}{n^2}$, $h_{2}^{(HA)} = \frac{-B^{(HA)}}{(n +
2)^2}$, $h_{3}^{(HA)} = \frac{-B^{(HA)}}{(n + 1)^2}$. To have solutions not
so complicated in form and more legible we take $\alpha = 0$.
Otherwise the calculations become very complicated. We find $\varrho_3^{(HA)} =
\frac{1}{3}$, $\varrho_1^{(HA)} = \frac{(n + 2)^2(2n + 1) -
\frac{1}{3}(6n^3 + 21n^2 + 24n + 8) }{6n^2 + 12n + 4}$,
$\varrho_2^{(HA)} = \frac{\frac{1}{3}(6n^3 + 15n^2 + 12n + 4) -
2n^3 - 3n^2 }{6n^2 + 12n + 4}$ and $|\varrho^{(HA)} |^2 =
\frac{\beta^2(B^{(HA)})^2(4n^2 + 8n + 3)}{n^2(n + 2)^2(n + 1)^4}$. Then for
both Hamiltonians we obtain the same type of solution
\begin {equation}
\rho_{int}[1](t) = \left(\begin{array}{ccc}
 1/3 +  \zeta (t) & 0 & \xi (t) \cr
0 & 1/3 - \zeta (t)& \overline{\xi (t)} \cr
 \overline{\xi (t)} &   \xi (t) & 1/3\cr
\end{array}\right).
\end {equation}
For HO we have
\be
 \zeta (t) &=& |\varrho'|(1 - \frac{4}{2 + |D|^2e^{\frac{2|\varrho'
 |^2t}{\beta}}}) \cr &=& |\varrho'|\tanh(\frac{|\varrho'
 |^2t}{{\beta}} + \ln(\frac{|D|}{\sqrt{2}})) \equiv  |\varrho'|\tanh(\theta' t + \vartheta')\cr
 \xi (t) &=& -\frac{2|\varrho'|D(1 + i)e^{\frac{|\varrho'
 |^2t}{\beta}}}{2 + |D|^2e^{\frac{2|\varrho'
 |^2t}{\beta}}}\cr &=& -\frac{D}{\sqrt{2}|D|}|\varrho'|{\rm sech}(\frac{|\varrho'
 |^2t}{{\beta}} + \ln(\frac{|D|}{\sqrt{2}})) \equiv Z'{\rm sech}(\theta' t + \vartheta')
\ee

and for HA

 \be
 \zeta (t) &=& |\varrho^{(HA)}|(1 - \frac{4d(n)}{2d(n) +
 |D|^2e^{\frac{2|\varrho^{(HA)}
 |^2t}{\beta}}})\cr &=&  |\varrho^{(HA)}|\tanh(\frac{|\varrho^{(HA)}
 |^2t}{{\beta}} + \ln(\frac{|D|}{\sqrt{2d(n)}})) \equiv |\varrho^{(HA)}|\tanh(\theta^{(HA)} t + \vartheta^{(HA)})\cr
 \xi (t) &=& -\frac{2|\varrho^{(HA)}|e(n)De^{\frac{|\varrho^{(HA)}
 |^2t}{\beta}}}{2d(n) + |D|^2e^{\frac{2|\varrho^{(HA)}
 |^2t}{\beta}}} = -\frac{De(n)}{\sqrt{2d(n)}|D|}|\varrho^{(HA)}|{\rm sech}(\frac{|\varrho^{(HA)}
 |^2t}{{\beta}} + \ln(\frac{|D|}{\sqrt{2d(n)}}))\cr &\equiv& Z^{(HA)}{\rm sech}(\theta^{(HA)} t + \vartheta^{(HA)}),
\ee where $d(n) = 4(2n + 1)(n+1)^3$, $e(n) = (2n + 1)(n + 2) + i
n\sqrt{(2n + 3)(2n + 1)}$ and $D =
\frac{\overline{\gamma_3}\gamma_1}{|\gamma_1|^2}$. Rewriting
$\rho_{int}[1](t)$ in spectral basis of the Hamiltonian and
renumerating in the order of increasing eigenvalues we find:
\begin {equation}
\rho_{int}[1](t) = \left(\begin{array}{ccc}
 1/3   &  \sqrt{2}\Re(\xi (t)) & \zeta (t) \cr
\sqrt{2}\Re(\xi (t)) & 1/3 & -i\sqrt{2}\Im(\xi (t)) \cr
 {\zeta (t)} & i\sqrt{2}\Im(\xi (t)) & 1/3\cr
\end{array}\right).
\end {equation}
If we compare this expression with formula (16) in \cite{CDSW} we
can see that we have two differences. First, the diagonal elements
are different but in both cases constant. The second difference is
more important. (16) describes $\rho[1](t)$ but not
$\rho_{int}[1](t)$ and we have to sandwich the latter between
$e^{i\alpha (\lambda )H t}$ and its inverse in sense of
(\ref{sandw}) to obtain the former. After this operation the
symmetry of $\rho[1](t)$ for HO and HA will be slightly different.
For HO we have exactly the same symmetry as in \cite{CDSW} and for
HA $\xi$ in the first row is multiplied by $e^{i\alpha (\lambda )(
h_1 - h_2 ) t}$ and in second row by $e^{i\alpha (\lambda )( h_2 -
h_3 ) t}$ (for HO both are the same).

Obviously these differences do not disturb the idea of \cite{CDSW}
to construct a time dependent Hamiltonian describing a three-level
perturbation of not only HO but for any system, e.g. HA
interacting with an optical soliton. The only modification comes
from non-equal spacings between levels of the system and produces
appropriate frequencies of oscillation. We can easily observe that
the only modification of $H$ is in multiplication by the factor
$\frac{2}{3}$ that, for the quadratic case, is equal exactly to
$\alpha (\lambda )$. Therefore we can start with Hamiltonian times
$\frac{3}{2}$ and get an appropriate $H_o$.

The relation $i\dot \rho=[H,\rho^2] =[H\rho+\rho H,\rho] =
[h,\rho]$ is given up to two parameters $\varepsilon_1$ and
$\varepsilon_2$: $h = (H + {\varepsilon}_1)\rho + (H +
{\varepsilon}_1)\rho + {\varepsilon}_2 1$. Putting
${\varepsilon}_1 = -\frac{h_1 + h_3}{2}$ and ${\varepsilon}_2 =
\frac{3(h_1 + h_3)}{4}$ we get:
\begin {equation}
h = \left(\begin{array}{ccc}
 2/3 h_1   &  0 & 0 \cr
0 & 2/3 h_2 & 0 \cr
 0 & 0 & 2/3 h_3\cr
\end{array}\right)
 + \left(\begin{array}{ccc}
 0   &  a(t) & 0 \cr
\overline{a(t)}& 0 & b(t) \cr
 0 & \overline{b(t)}& 0\cr
\end{array}\right)=H_o + V = H_o -\vec{d}\cdot \vec{E},
\end {equation}
where $a(t) = (h_2 - h_3)\sqrt{2}\Re(\xi (t))e^{i 2/3 (h_1 - h_2 )
t}$ and $b(t) = -i(h_2 - h_1)\sqrt{2}\Im(\xi (t))e^{i 2/3 (h_3 -
h_2 ) t}$. From this we can see that:

\be \vec{E} = (E_x ,E_y ,E_z) = (E_x (0){\rm sech}(\theta t +
\vartheta)\cos(\omega t), E_y (0){\rm sech}(\theta t +
\vartheta)\sin(\omega t), 0) \label{e}
\ee

and

\begin {equation}
d_x = \left(\begin{array}{ccc}
 0   &  \tilde{a}(t)\cos(\omega t) & 0 \cr
\overline{\tilde{a}(t)}\cos(\omega t)& 0 & \tilde{b}(t)\cos(\omega
t) \cr
  0& \overline{\tilde{b}(t)}\cos(\omega t)& 0\cr
\end{array}\right),
 d_y = \left(\begin{array}{ccc}
 0   &  \tilde{a}(t)\sin(\omega t) & 0 \cr
\overline{\tilde{a}(t)}\sin(\omega t)& 0 & \tilde{b}(t)\sin(\omega
t) \cr
 0 & \overline{\tilde{b}(t)}\sin(\omega t)& 0\cr
\end{array}\right)
\end {equation}
$\tilde{a}(t) = (h_2 - h_3)\sqrt{2}\Re(Z)e^{i 2/3 (h_1 - h_2 )
t}$,  $\tilde{b}(t) = -i(h_2 - h_1)\sqrt{2}\Im(Z)e^{i 2/3 (h_3 -
h_2 ) t}$. Hence we get potential $V$ in shape of well-known
McCall-Hahn ``sech" soliton \cite{McCH}. Another choice of
parameters ${\varepsilon}_i$ leads us to a very wide class of
electromagnetic impulses \cite{I}. We can use now this solution to
construct another one for the same potential in accordance with
the SUSY scheme. The possibility follows from the richness of the
set of solutions we obtain by the Darboux transformation.

Now we are in position to check that we have not only the solution
of the Bloch equation but also of the Maxwell one. First we
calculate the components of atomic polarization:

\be
 P_x = {\rm Tr} (\rho d_x) = 4{\rm sech}(\theta t +
\vartheta)\cos(\omega t)[(h_2 - h_3)\Re(Z)^2 + (h_2 -
h_1)\Im(Z)^2] \cr
 P_y = {\rm Tr} (\rho d_y) = 4{\rm sech}(\theta t +
\vartheta)\sin(\omega t)[(h_2 - h_3)\Re(Z)^2 + (h_2 -
h_1)\Im(Z)^2] \label{p} \ee

Let us remember that $(h_2 - h_3)$ and $(h_2 - h_1)$ have the same
sign.

Because McCall-Hahn is a steady-state pulse we can use coordinate
$\varsigma = t - \frac{z}{v}$, where $v$ is the constant pulse
velocity. Hence we have $\frac{\partial}{\partial t}\mapsto
\frac{d}{d\varsigma}$ and $\frac{\partial}{\partial z}\mapsto
-\frac{1}{v}\frac{d}{d\varsigma}$ then Maxwell equation has the
following shape:
$$\frac{d^2}{d\varsigma^2}\vec{E} = (\frac{4\pi}{(\frac{c}{v})^2 -
1})\frac{d^2}{d\varsigma^2}\vec{P}.$$

By comparing (\ref{e}) and (\ref{p}) we can see that Maxwell
equation is fulfilled if we choose parameters in such a way:
$$16\pi[(h_2 - h_3)\Re(Z)^2 + (h_2 -
h_1)\Im(Z)^2] = ((\frac{c}{v})^2 - 1)E_o (0),$$ where $o$ means
$x$ or $y$.

\section{Comments}

We have described a construction of solutions for the two main
``building blocks": $2\times 2$ and $3\times 3$. How to get from
$2\times 2$ blocks to infinite-dimensional and irreducible
solutions was discussed in \cite{UCKL}. However, in infinite
dimensional examples the issues of normalization to $\Tr \rho=1$
involve certain subtleties that require a separate treatment.
These difficulties do not occur if one restricts the discussion to
arbitrarily large but finite matrix dimensions.  The method we
have described in the present paper allows for immediate
generalizations to arbitrary finite dimensions if one follows the
strategy employed in \cite{UCKL}.

The technique gives a method of constructing three-dimensional
solutions appropriate for different situations. It has to be
stressed that not only can we use any Hamiltonian and any type of
nonlinearity, but it seems that we can get arbitrary values for
the eigenvalues of $\rho$ for a homogeneous nonlinearity. To see
this it is enough to observe that we can generalize the scaling
and the shifting properties which were observed in \cite{LC} the
for quadratic nonlinearity. If $\varrho$ is solution of NvNE for
$f(x) = x^k$ then $\frac{\rho}{Tr \rho}(\frac{t}{(Tr
\rho)^{k-1}})$ also satisfies it and $\rho + \sigma$ when
$\dot{\sigma}= 0$ is solution if we put $g(x) = f(x + \sigma )$
instead of $f(x)$. Because $\varrho_1$ and $\varrho_2$ are very
simple functions of the parameter $\alpha$ and $\varrho_3$, and
$\varrho$ depends only on $\beta$, we can produce positive
``predensity" matrices with wide spectrum of eigenvectors. How
``big" is this set of solutions in the set of all the solutions of
NvNE will be considered elsewhere.

\acknowledgments

This work was started as a part of the KBN Project No. 5P03B 040
20,  and the Flemish Research Fund (FWO) Project G.0359.03
``Soliton concept in classical and quantum contexts". The paper
was completed with the support of the KBN Grant PBZ-Min-008/P03/03
and the Flemish-Polish bilateral collaboration ``Soliton
techniques applied to equations of quantum field theory". I am
indebted to Jan Naudts and Marek Czachor for comments.

\end{document}